\documentclass{IEEEtran4PSCC}

\ifCLASSINFOpdf
   \usepackage[pdftex]{graphicx}
\else
   \usepackage[dvips]{graphicx}
\fi

\usepackage{multirow}
\usepackage[cmex10]{amsmath}
\usepackage{amsfonts,amsthm}
\usepackage{makecell}
\usepackage{rotating}
\usepackage{xcolor}
\usepackage{multicol}
\usepackage{multirow}
\usepackage{booktabs}
\usepackage{enumitem}
\usepackage{longtable}
\usepackage{fancyhdr}
\newtheorem{theorem}{Theorem}

\newtheorem{remark}{Remark}

\newtheorem{assumption}{Assumption}

\makeatletter
\let\old@ps@headings\ps@headings
\let\old@ps@IEEEtitlepagestyle\ps@IEEEtitlepagestyle
\def\psccfooter#1{%
    \def\ps@headings{%
        \old@ps@headings%
        \def\@oddfoot{\strut\hfill#1\hfill\strut}%
        \def\@evenfoot{\strut\hfill#1\hfill\strut}%
    }%
    \def\ps@IEEEtitlepagestyle{%
        \old@ps@IEEEtitlepagestyle%
        \def\@oddfoot{\strut\hfill#1\hfill\strut}%
        \def\@evenfoot{\strut\hfill#1\hfill\strut}%
    }%
    \ps@headings%
}
\makeatother

\psccfooter{%
        \parbox{\textwidth}{\hrulefill \\ \small{23rd Power Systems Computation Conference} \hfill \begin{minipage}{0.2\textwidth}\centering \vspace*{4pt} \includegraphics[scale=0.06]{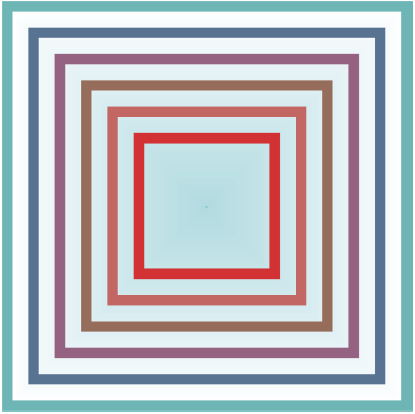}\\\small{PSCC 2024} \end{minipage} \hfill \small{Paris, France --- June 4 -- 7, 2024}}%
}

\begin{document}


\title{Decomposed Phase Analysis using Convex Inner Approximations: a Methodology for DER Hosting Capacity in Distribution Systems}

\author{
\IEEEauthorblockN{Hani Mavalizadeh, Mads R. Almassalkhi}
\IEEEauthorblockA{Department of Electrical and Biomedical Engineering\\ 
University of Vermont\\
Burlington, USA\\
\{hmavaliz, malmassa\}@uvm.edu}
}


\maketitle
\begin{abstract}
This paper uses convex inner approximations (CIA) of the AC power flow to tackle the optimization problem of quantifying a three-phase distribution feeder's capacity to host distributed energy resources (DERs). This is often connoted hosting capacity (HC), but herein we consider separative bounds for each node on positive and negative DER injections, which ensures that injections within these nodal limits satisfy feeder voltage and current limits and across nodes sum up to the feeder HC. The methodology decomposes a three-phase feeder into separate phases and applies CIA-based techniques to each phase. An analysis is developed to determine the technical condition under which this per-phase approach can still guarantee three-phase constraints. New approaches are then presented that  modify the per-phase optimization problems to overcome conservativeness inherent to CIA methods and increase HC, including selectively modifying the per-phase impedances and iteratively relaxing per-phase voltage bounds. Discussion is included on trade-offs and feasibility. To validate the methodology simulation-based analysis is conducted with the IEEE 37-node test feeder and a real 534-node unbalanced radial distribution feeder.
\end{abstract}


\thanksto{\noindent The authors graciously recognize support from the National Science Foundation (NSF) Award ECCS-2047306.}

\section{Introduction}
As the deployment of distributed energy resources (DERs) in power grids continues to accelerate, their utilization in a number of ancillary services is increasing~\cite{Schmitt2023MPSCE}. In this context, DERs can be managed by aggregators, which dispatch them in response to market signals, often without taking into account the limitations of the grid. This lack of consideration can potentially lead to violations of critical grid constraints, including voltage and transformer limits. Therefore, there is an urgent need for what is referred to as \emph{Grid-aware DER coordination}, which involves effectively accounting for AC network constraints during the coordination of DERs~\cite{Almassalkhi2020Energies}.

Various methods have been proposed in the technical literature for grid-aware DER coordination. One common approach is to restrict the amount of power that each customer can export to the grid~\cite{AusNetServices2017}. However, this method can be overly conservative, and with the rapid increase in the number of DERs connected to the grid, these fixed limits can become outdated and require frequent updates~\cite{Ochoa2022TSG}.

In direct control schemes, it is assumed that the grid operator has access to all DER data and can directly control DERs~\cite{Chen2020TSG, Dallanese2018TPWRS}. While direct control methods can theoretically provide optimal solutions, they often rely on strong assumptions related to observability and controllability. In practice, DER aggregators do not have access to grid data, and grid operators do not have full control over DERs.

Alternatively,~\cite{hesamzadeh2018PSCC} proposes an approach where the grid operator adjusts locational marginal prices (LMPs) based on grid conditions to incentivize the aggregator to adapt the DER aggregate load accordingly. However, this paper assumes a balanced distribution system, which may not hold in real-world applications. In~\cite{mathieu2021TPWRS}, two mechanisms are presented to allow the grid operator to override DER aggregator dispatch decisions to ensure grid constraints are not violated. One limitation is that in certain electric markets, the grid operator may not have the authority to block aggregator control decisions.

Another approach is for the grid operator to establish limits on the amount of injection from each node to preserve grid constraints. This approach requires minimal information exchange between the grid operator and aggregator. In~\cite{Ochoa2022TSG}, the concept of \emph{operating envelopes} is introduced, where the grid operator uses linear or model-free methods to issue time-varying export/import limits to aggregators. A convex inner approximation (CIA) is presented in~\cite{nazir2021TPWRS} for maximizing voltage margins, which is generalized in~\cite{Nawaf2022TPWRS} to compute feeder hosing capacity of balanced or single-phase distribution feeders.  In~\cite{lee2020TPWRS}, a sequential algorithm is presented that constructs a convex restriction around an initial feasible point, subsequently refining it to obtain an improved feasible solution. This work is extended further in~\cite{leedong2021TPWRS}, where the approach is enhanced to account for robustness against uncertainty in power injections. In~\cite{bassi2022Report}, a model-free approach is introduced, leveraging historical meter data and neural networks to eliminate the need for solving the non-convex AC OPF problem in unbalanced distribution feeders. It demands access to substantial volumes of meter data, which may not always be readily available. Additionally, it's important to note that model-free methods can exhibit sensitivity to the quality and distribution of data. In~\cite{Petrou2021TSG} a bottom-up approach is presented where DERs submit power injection requests based on their local controllers to the grid operator. The grid operator can deny injection requests if a three-phase power flow analysis indicates a risk of grid constraint violation. An optimization model for assessing the hosting capacity (HC) of DERs, taking into consideration the anticipated network conditions during demand response scheduling and adapting to the real-time network state is developed in~\cite{Rigoni2021TPWRS}. 

Thus, in the literature, there are either simplified models used to compute hosting capacities with no guarantees or guarantees applicable only to simplified systems. It is within this context that this paper contributes to the field of computing hosting capacity for realistic systems with outlined trade-offs between optimality and guaranteed feasibility:
\begin{itemize}
\item The recently presented optimization-based approach for computing the (dynamic) hosting capacity of single-phase distribution feeders in~\cite{Nawaf2022TPWRS} has been extended to three-phase, unbalanced distribution feeders. This extension is further adapted to account for mutual impedances in the original optimization problem. 

\item The HC estimate for unbalanced feeders is then improved by iteratively adjusting voltage bounds within the per-phase optimization framework, accounting for mutual impedances and unbalanced load in the 3-phase system. 
\item Finally, the methodology is validated through simulation-based analysis on the IEEE 37-node feeder and a real 3-phase network with more than 500 three-phase nodes.
\end{itemize}

The remainder of the paper is organized as follows: Section~\ref{sec:HC} provides a concise overview of the preliminary concepts used in this paper. The proposed approach to extend the CIA-based method to three-phase unbalanced grids is presented in Section~\ref{sec:extension}. In Section~\ref{sec:guarantee}, an approach for modifying the optimization problem based on the 3-phase feeder data is detailed. Section~\ref{sec:DHC} introduces an iterative method designed to enhance HC. Finally, numerical results are provided in Section~\ref{sec:numerical} followed by concluding remarks in Section~\ref{sec:conclusion}.

\section{Preliminaries}\label{sec:HC}

This section describes the modeling of distribution feeders and their hosting capacity (HC). The CIA-based approach detailed in~\cite{nazir2021TPWRS,Nawaf2022TPWRS} employs a CIA of the set of feasible admissible injections. An optimization problem is used to determine the HC at each node of a distribution feeder. The total feeder HC is the sum across all nodes. We aim to adapt this for unbalanced feeders. Next, we discuss the single-phase equivalent load flow, i.e.,~\textit{DistFlow}~\cite{baran_optimal_1989}, and the convex HC formulation.

\subsection{Balanced feeder HC via convex inner approximations}

Consider a radial (single-phase) distribution feeder as a tree graph $G = (\mathcal{V}, \mathcal{E})$ with $N$ nodes $\mathcal{V} := \{1, \ldots, N \}$ and $N-1$ branches $\mathcal{E} \subseteq \mathcal{V} \times \mathcal{V}$, such that if nodes $i$ and $j$ are connected, then $(i, j) \in \mathcal{E}$. At each node $i\in\mathcal{V}$, \textit{DistFlow} considers the square of the voltage phasor magnitude, i.e., $V_i:=|v_i|^2$ and complex power injections are denoted $s_i = p_i + \mathbf{j}q_i$. Node~0 is assumed to be the substation (slack) node with a fixed voltage $V_0$. Through each branch with impedance $z_{ij} = r_{ij} + \mathbf{j} x_{ij}$, we consider the square of the current phasor, i.e., $l_{ij} := |I_{ij}|^2$ and the active and reactive power flows, $P_{ij}$ and $Q_{ij}$.
Thus, from the \textit{DistFlow} formulation and applying~\cite{heidari2017ANZCC}, the relationships between voltages and branch power flows and nodal injections and line currents can be defined as
\begin{subequations} \label{eq:DistFlowLin}	
\begin{align}
    V &= V_0\mathbf{1}_N + M_p p + M_q q - Hl \label{eq:V-P}\\
    P &= Cp - D_R l\\
    Q &= Cq - D_X l,
\end{align}
\end{subequations}
where appropriately-sized matrices $M_p$, $M_q$, $H$, $C$, $D_R$, $D_X$ are detailed in~\cite{nazir2021TPWRS, Nawaf2022TPWRS} and serve to map injections and currents to corresponding voltages and branch power flows across the network. Besides the linear equations in~\eqref{eq:DistFlowLin}, the \textit{DistFlow} also relates voltages and power flows to currents via non-convex
\begin{align}\label{eq:current}
    l_{ij}(P,Q,V) = (P_{ij}^2 + Q_{ij}^2)/V_i.
\end{align}
The non-linear~\eqref{eq:current} makes the \textit{DistFlow} formulation non-convex within an optimal power flow (OPF) setting. Thus, we are interested in utilizing a CIA of the \textit{DistFlow} formulation. The CIA effectively bounds the nonlinear $l_{ij}$ with a convex envelope: $l^-(P,Q,V) \le l_{ij}(P,Q,V) \le l^+(P,Q,V)$, which enables the creation of two sets of variables: upper ($^+$) and lower proxies ($^-$), e.g.,  $V^-\le V \le V^+$. As long as the lower proxies satisfy lower limits and upper proxies satisfy upper limits, e.g., $\underline{V}\le V^-$ and $V^+ \le \bar{V}$, then we are guaranteed that the physical variable satisfies, e.g., $\underline{V} \le V \le \bar{V}$. This guarantee means that we can replace the physical variables altogether and replace them with their convex proxies.



Consider for example a feeder with inductive branches, i.e., $x_{ij} > 0, \,\, \forall (i,j)\in\mathcal{E}$~\cite{nazir2021TPWRS}. Then, we can replace the non-convex formulation in~\eqref{eq:DistFlowLin} and~\eqref{eq:current} with their convex proxies: 
\begin{subequations}\label{eq:proxies}
\begin{align}
V^+ &= V_0\mathbf{1}_N + M_pp + M_qq - Hl^-\\
V^- &= V_0\mathbf{1}_N + M_pp + M_qq - Hl^+\\
P^+ &= Cp - D_R l^{-}\\
P^- &= Cp - D_R l^{+}\\
Q^+ &= Cq - D_{X} l^{-}\\
Q^- &= Cq - D_{X} l^{+}.\\
l^+ &\ge  f_\text{quad}(P^{+},P^{-},Q^{+},Q^{-},V^+,V^-) \label{eq:currProxyUpper}\\
l^- &:= f_\text{aff}(P^{+},P^{-},Q^{+},Q^{-},V^+,V^-), \label{eq:currProxyLower}
\end{align}
\end{subequations}
where $l^-$ is affine in the proxy variables while $l^+$ is a convex relaxation of a quadratic function of the proxy variables. Please see Appendix~\ref{sec:appx} for derivations of $f_\text{aff}$ and $f_\text{quad}$ and~\cite{nazir2021TPWRS,Nawaf2022TPWRS} for full details. Finally, the feeder HC is then the maximum sum of nodal injections, $p_i^+:= p_i^\ast$, that drives the feeder to its capacity (e.g., voltage, current, or power flow limits are active). The convex formulation that achieves this objective is 
\begin{subequations} \label{eq:CIAproxy}
\begin{align}
\mathbf{P}_\text{CIA}^{\phi,+}: \quad  p^+ := \arg\max_{p_i}& \sum_{i=1}^N w_i p_{i}\\
    \text{subject to} & \,\, \eqref{eq:proxies}\\
    &\underline{l} \le l^-\qquad   l^+ \le \overline{l} \\ 
    &\underline{V} \le V^-\quad V^+\le \overline{V} \\
    & p_i^2+q_i^2\le \overline{s_i}^2,\quad \forall i\in \mathcal{V}, \label{eq:sbound}
\end{align}
\end{subequations}
where $w_i$ are design parameters that differentiate nodal capacities. Note that inequality~\eqref{eq:sbound} is optional and captures limits on active injections based on apparent power limits at each node (e.g., from inverter, transformer, or power factor limits). Other constraints on $P^{+/-},Q^{+/-}$ may also be added.

The HC for DER injections (e.g., solar PV) is defined as
\begin{align}
 \overline{\text{HC}} := \sum_{i}^N p_i^+ = \mathbf{1}_N^\top p^+ > 0.
\end{align}
 Similarly, we can define the HC relative to consumption (e.g., electric vehicle HC) as  $\underline{\text{HC}} := \sum_{i}^N p_i^- = \mathbf{1}_N^\top p^- < 0$, where $p^-$ is the solution that minimizes the nodal (net) injections, i.e., solve corresponding $\mathbf{P}_\text{CIA}^{\phi,-}$ problem, whose details are omitted due to page limits. Thus, $\underline{\text{HC}} \le 0 \le \overline{\text{HC}}$.

However, since $\mathbf{P}_\text{CIA}^{\phi,+}$ and $\mathbf{P}_\text{CIA}^{\phi,-}$ employ a CIA of \textit{DistFlow}, the HC estimates are valid only for balanced, radial distribution feeders. We are now interested in how to adapt this CIA-based method to a realistic unbalanced distribution feeder, which means that we need to consider the effects of mutual phase impedance and load unbalances. 





\subsection{Unbalanced distribution power flow}

Consider a three-phase, radial graph $G$, wherein each node represents three phases: $a$, $b$, $c$. Similarly, each branch represents a three-phase line section with a corresponding $3\times 3$ impedance matrix, which is expressed as,
\begin{align}\label{eq:z_line}
    z_{ij}^{3\phi} := \begin{bmatrix}
    z^{\text{a}}_{ij} & z^{\text{ab}}_{ij}  & z^{\text{ac}}_{ij}  \\
    z^{\text{ba}}_{ij}  & z^{\text{b}}_{ij}  & z^{\text{bc}}_{ij}  \\
    z^{\text{ca}}_{ij}  & z^{\text{cb}}_{ij}  & z^{\text{c}}_{ij}  \\
    \end{bmatrix}\forall (i,j)\in \mathcal{E}.
\end{align}



Voltage at 3-phase node $i$ is denoted $V_{i}^{3\phi}=\begin{bmatrix}
    V_{i}^a,V_{i}^b,V_{i}^c
\end{bmatrix}^\top$ and current in branch $(i,j)\in \mathcal{E}$ is $I_{ij}^{3\phi}=\begin{bmatrix}
    I_{ij}^a,I_{ij}^b,I_{ij}^c
\end{bmatrix}^\top$. The line voltage drop and currents are related by
\begin{align}
    \Delta V^{3\phi}_{ij} := V^{3\phi}_{i} - V^{3\phi}_{j}   = z_{ij}^{3\phi}I^{3\phi}_{ij} \Rightarrow\Delta V^{3\phi} = Z^{3\phi} I^{3\phi}, \label{eq:volt_drop}
\end{align}
where, $I^{3\phi}=[I_{ij}^{3\phi}]_{(i,j)\in \mathcal{E}} \in \mathbb{C}^{3(N-1)}$ represents the complex three-phase currents, $V^{3\phi}=[V_{i}^{3\phi}]_{i\in \mathcal{V}}\in \mathbb{C}^{3N}$ corresponds to the three-phase voltages, $Z^{3\phi} \in \mathbb{C}^{3(N-1) \times 3(N-1)}$ is the complex three-phase impedance matrix

Next, we seek to extend the CIA-based method from balanced (single-phase equivalent) feeders to unbalanced feeders.

\section{Extending CIA to unbalanced feeders}\label{sec:extension}
Given an unbalanced feeder, how can we approximate or decompose it for HC analysis? In this section, we seek to answer this question. Specifically, we consider methods for 1) approximating feeders as balanced (e.g., by modifying line impedances and nodal loads and 2) decomposing feeders along their phases. These are summarized next. 

\begin{itemize}
    \item \textbf{Method 1 - balanced feeder approximation:} This strategy involves transforming an unbalanced feeder into an \textit{approximate} balanced model, which is then used to determine $p^-, p^+$ from $\mathbf{P}_\text{CIA}^{\phi,+\text{\textbackslash} -}$. The resulting per-phase HC is then distributed equally to each phase. We consider two different ways to approximate a balanced feeder:
    \begin{enumerate}
        \item[i)] Take the maximum line impedance and minimum loads across all three phases to capture the worst-case voltage drop/rise. 
        \item[ii)] Average line impedances and loads across phases a/b/c to create a balanced approximation of a feeder. This approximation can potentially cause voltage violations at the corresponding HC value. 
    \end{enumerate}
    
    \item \textbf{Method 2 - per-phase analysis:} In this approach, we extract each phase separately and compute $p^-$ and $p^+$. This per-phase approach is considered for two different implementations:
    \begin{enumerate}
        \item[i)] One phase is selected and nodal HC values, $(p_i^-,p_i^+)$, are computed for that phase. For the 3-phase feeder, the same $(p_i^-,p_i^+)$ values are then applied to all phases at a three-phase node. We denote the sub-methods $2\text{i}^\phi$ for $\phi = \{\text{a},\text{b},\text{c}\}$, e.g., $\overline{\text{HC}}_{3\phi} = 3 \times \mathbf{1}^\top_N p^+_{\text{a}}$ for method $2\text{i}^a$.
        \item[ii)] All three phases are extracted separately and we compute  $(p^-,p^+)$ for each phase, which yields hosting capacity, e.g., $\overline{\text{HC}}_{3\phi} = \mathbf{1}_N^\top (p^+_{\text{a}} + p^+_{\text{b}} + p^+_{\text{c}})$.
    \end{enumerate}
    
    
\end{itemize}

\begin{figure}
    \centering
    \includegraphics[width=1\columnwidth]{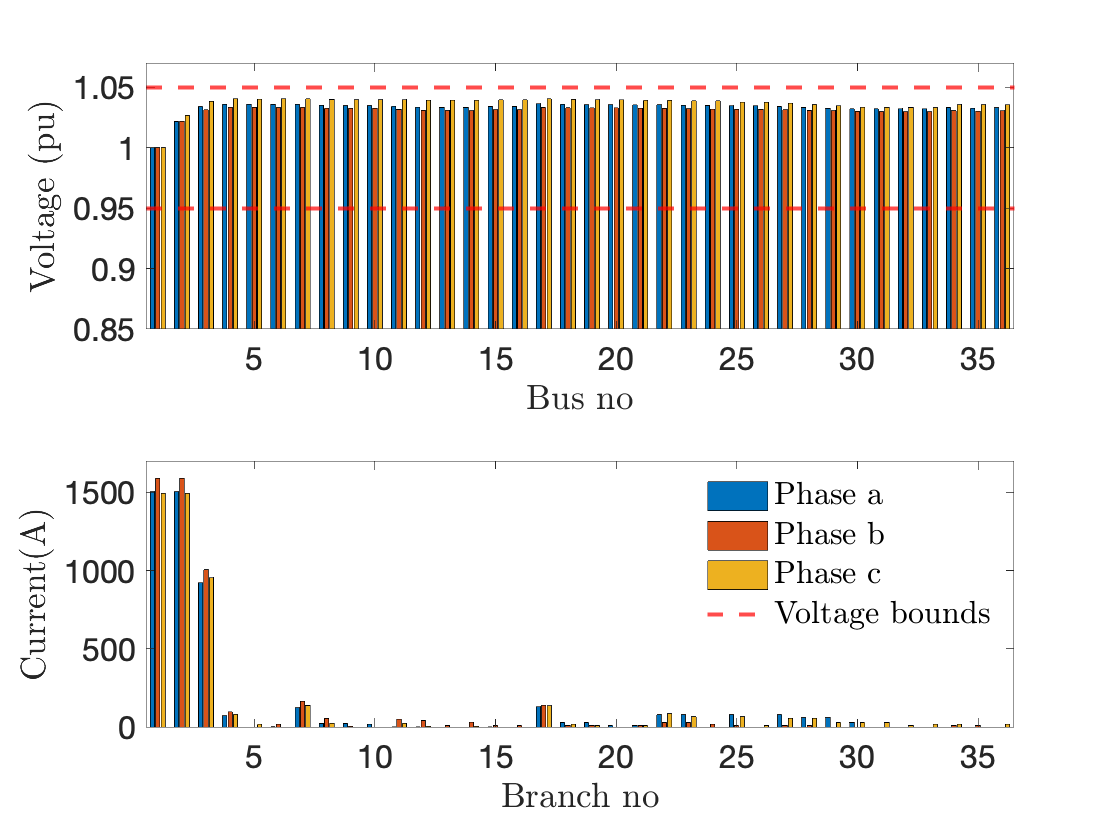}
    \caption{Illustrating the effects of Method 2ii on three-phase voltage and current profiles following the addition of nodal injections $p^+_a + p^+_b + p^+_c$. The dashed red line indicates the ANSI voltage limits of [0.95, 1.05]~pu.}
    \label{fig:HC}
\end{figure}

Each of these methods estimates the three-phase HC, e.g., $\overline{\text{HC}}$ by computing net nodal injections, e.g., $p^+_{\phi}$, which are then applied to the full three-phase network to determine the corresponding three-phase voltages and currents. In Fig.\ref{fig:HC}, these voltage and current profiles for Method 2ii are presented for the IEEE 37-node test feeder~\cite{schneider2018ITPWRS}. As can be seen, despite single-phase analysis underpinning the HC estimate, phase voltages are within $\overline{V}=1.05$~pu across all nodes and phases. Next, we are interested in metrics that can be used to compare the different methods.

\begin{itemize}
\item \textbf{Total number of violations}, $N_{v}$, counts the number of nodes and phases for which $|V_{i}^{3\phi}| \notin [\underline{V}, \overline{V}]$.

    
    \item \textbf{Maximum violation in per unit}, $M_v$, provides a measure of the severity of the violations:
        \begin{align}
        M_v = \max_{i=1,\hdots, 3N}\left\{ \max\left\{0, E^{\text{u}}_i, E^{\text{l}}_i\right\}\right\},
    \end{align}
    where, $E^{\text{u}}:=|V^{3\phi}| - \overline{V}\mathbf{1}_{3N}$ and $E^{\text{l}}:=\underline{V}\mathbf{1}_{3N} - |V^{3\phi}|$.

    \item \textbf{Sum of violations}, $S_v$, captures the cumulative severity of violations across the network:
        \begin{align}
        S_v =  \sum_{i=1}^{3N}\max\left(0, E^{\text{u}}_i, E^{\text{l}}_i\right).
    \end{align}
    
    \item \textbf{Average voltage margin}, $W_M$, measures how conservative the HC results from $\mathbf{P}_\text{CIA}^{\phi,+/-}$ are:
    \begin{align}
        W_M = \frac{1}{3N} \sum_{i=1}^{3N} \max\left\{0,\Delta W_i\right\},
    \end{align}
    where $\Delta W_i:=\min\left\{|V_i^{3\phi}| - \underline{V}, \overline{V} - |V_i^{3\phi}|\right\}$.
    
    \item \textbf{Voltage unbalance factor (VUF)} provides a relative measure (in \%) of voltage unbalance caused by nodal HC injections:
    \begin{align}
        \text{VUF} = \frac{100}{N} \sum_{i=1}^{N} \frac{\max\left \{|V_{i}^{3\phi}| - \frac{1}{3} \mathbf{1}_3^\top|V_{i}^{3\phi}|\mathbf{1}_{3}\right\}}{\frac{1}{3} \mathbf{1}_3^\top|V_{i}^{3\phi}|}.
    \end{align}

\end{itemize}

 It should be noted that none of the methods leads to voltage violation in the IEEE 37 node feeder. That is due to the inherent conservativeness of the CIA. 
To compare Methods~1 and~2, we consider three scenarios:
$i)$ Boost loads in phase~c by 20\% and decrease loads on phase~b by 20\%.
$ii)$ From scenario $i$, swap the loads of phases~b and~c.
$iii)$ From scenario $i$, swap loads of phases~a and~b. Throughout, the power factor is held constant.
Table~\ref{t:accuracy} uses the metrics above to compare minimum HC estimates, i.e., using $\mathbf{P}_\text{CIA}^{\phi,-}$, average of $W_M$ and VUF, sum of $S_v$ and $N_v$, across scenarios $i$, $ii$ and $iii$. Notably, the comparison shows that Method~2ii does not cause voltage violations. Method~1i leads to an overly loaded network which makes the optimization problem infeasible. Other methods result in voltage violations. Thus, based on results in Table~\ref{t:accuracy}, Method~2ii is selected for further analysis. Given that Method~2ii uses information from all phases without averaging, it was somewhat expected that Method~2ii could outperform the other approaches. It should be noted that considering the mutual impedance can lead to less or more conservative HC depending on the characteristics of $z^{m}_{ij}$.  In the next section, technical conditions are presented under which per-phase analysis and HC optimization extend to three-phase networks.

\begin{table}
\caption{Performance of the proposed methods}

\centering
\begin{tabular}{p{1.1cm} p{0.5cm} p{0.8cm} p{0.8cm} p{0.8cm} p{0.8cm} p{0.8cm}}
\toprule
        Method & \(N_v\) & \(M_v\)(pu)   & \(S_v\) (pu) & \(W_M\) (pu) &\(\underline{\text{HC}}\) (MW)& VUF (\%)\\
\midrule
1i & 0 & 0  & 0 & 0.039 & NA& 0.81 \\
1ii & 17 & 0.005  & 0.040 & 0.017 & -14.68& 0.87 \\
$2\text{i}^\text{a}$ & 18 & 0.010  & 0.085 & 0.017 & -14.03& 0.88 \\
$2\text{i}^\text{b}$ & 33 & 0.013  & 0.158 & 0.017 & -11.07 & 0.92\\
$2\text{i}^\text{c}$ & 4 & 0.003  & 0.009 & 0.022 & -10.03& 0.81 \\
2ii & 0 & 0  & 0 & 0.019 & -13.98& 0.48 \\
\bottomrule
\end{tabular}\label{t:accuracy}
\end{table}

\section{Modifying $\mathbf{P}_\text{CIA}^{\phi,-/+}$ for three-phase grid}\label{sec:guarantee}
In this section, we present an approach for adapting the per-phase HC estimates to deal with the inherent conservativeness of CIA. The method effectively modifies the impedance matrix to take into account the impact of mutual impedance. The approach makes the following assumptions:

\begin{assumption}\label{ass:balancedLoad}
The sum of the phase load currents is zero.     
\end{assumption}

\begin{assumption}\label{ass:imp}
Three-phase lines are transposed, such that mutual impedances are identical:
$z_{ij}^{3\phi}=
\begin{bmatrix}
z_{ij}^a  & z_{ij}^m    & z_{ij}^m    \\
z_{ij}^m    & z_{ij}^b  & z_{ij}^m     \\
z_{ij}^m   & z_{ij}^m     & z_{ij}^c
\end{bmatrix}
$.
\end{assumption}

From the above assumptions, the following theorem holds.

\begin{theorem}\label{thm:1}
Given a 3-phase system that satisfies Assumptions~\ref{ass:balancedLoad} and~\ref{ass:imp}, 
 if per-phase optimization $\mathbf{P}_\text{CIA}^{\phi, +}$ satisfies $\underline{V} \le V_i(p^+) \le \overline{V} \,\, \forall i\in \mathcal{V}$, then  the three-phase system satisfies $\underline{V} \le V_i^{3\phi}(p^+) \le \overline{V} \,\, \forall i\in \mathcal{V}$. Same holds for $\mathbf{P}_\text{CIA}^{\phi,-}$ and $V_i(p^-)$.
    
\noindent \textbf{Proof}: please see Appendix~\ref{appx:proof}.
\end{theorem}

Theorem~\ref{thm:1} states when a three-phase distribution feeder can be decomposed into three decoupled single-phase distribution systems with modified impedances, $z_{ij}^{\phi} - z_\text{ij}^m$, to provide guarantees that the resulting HC will not engender voltage violations in the three-phase system.

\begin{remark}
    Using a similar approach, and by further assuming identical conductor impedances $z_{ij}^a = z_{ij}^b = z_{ij}^c$, Theorem~\ref{thm:1} extends to Delta-connected loads. 
\end{remark}



In practical settings, when Assumptions~\ref{ass:balancedLoad} and~\ref{ass:imp} do not hold, $z_\text{ij}^m$ can be approximated by
\begin{align}\label{eq:zm_avg}
    z_\text{ij}^m \approx (z_\text{ij}^{\textbf{ab}}+z_\text{ij}^{\textbf{ac}}+z_\text{ij}^{\textbf{bc}})/3.
\end{align}
From each phase, we construct a sub-feeder from which we can compute nodal HC (net) injections $p^-_{i}$ and $p^+_{i}$ using Method~2ii. The resulting voltages of the single-phase networks, $|V_i|$, are then compared with those of the full 3-phase load flow, $|V_i^{3\phi}|$, with the 3-phase (net) injections $p_{3\phi}^+ := [p^+_{a,i},p^+_{b,i},p^+_{c,i}]_{i\in \mathcal{V}}$ added to the system load. We denote the approach of solving $\mathbf{P}_\text{CIA}^{+/-}$ with modified impedance from Theorem~\ref{thm:1} as \textbf{Mod-Z} HC.

\begin{figure}[t]
\centering
\includegraphics[width=1\columnwidth]{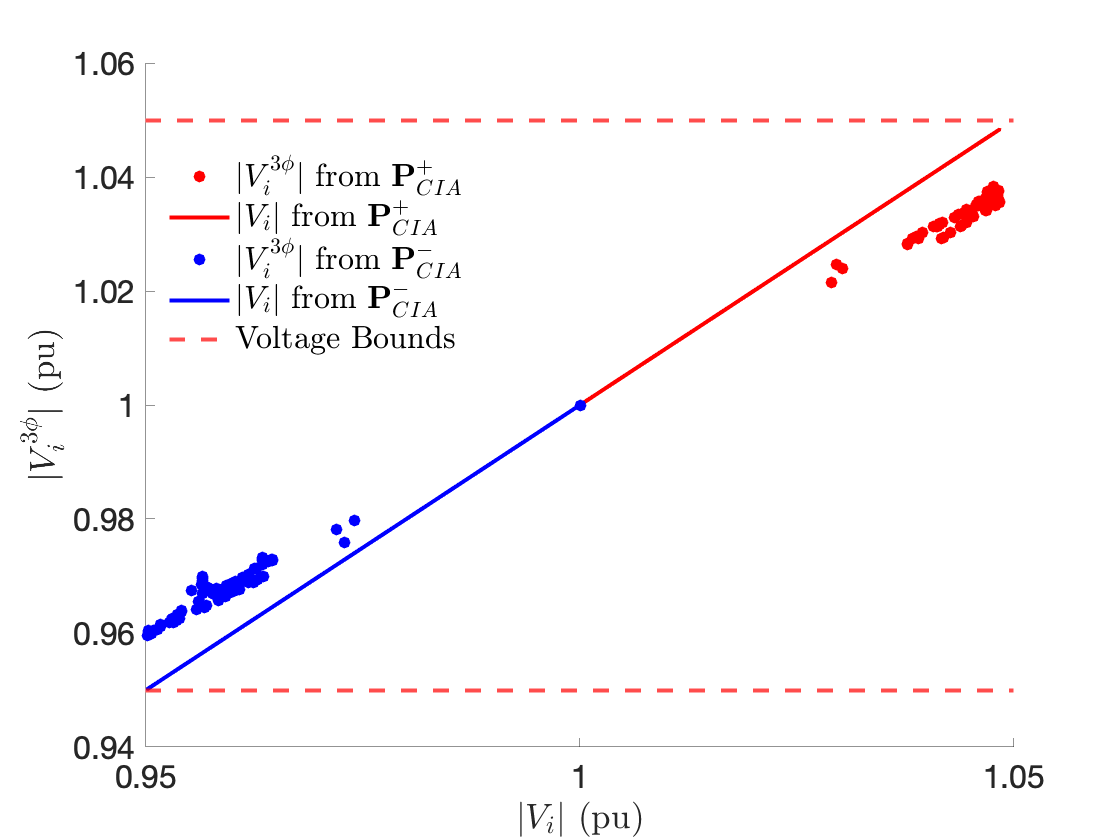}
 \caption{Comparison between three-phase and single-phase voltages for the modified IEEE 37-node system under Method~2ii.}
\label{fig:Base}
\end{figure}

\begin{figure}[t]
\centering
\includegraphics[width=1\columnwidth]{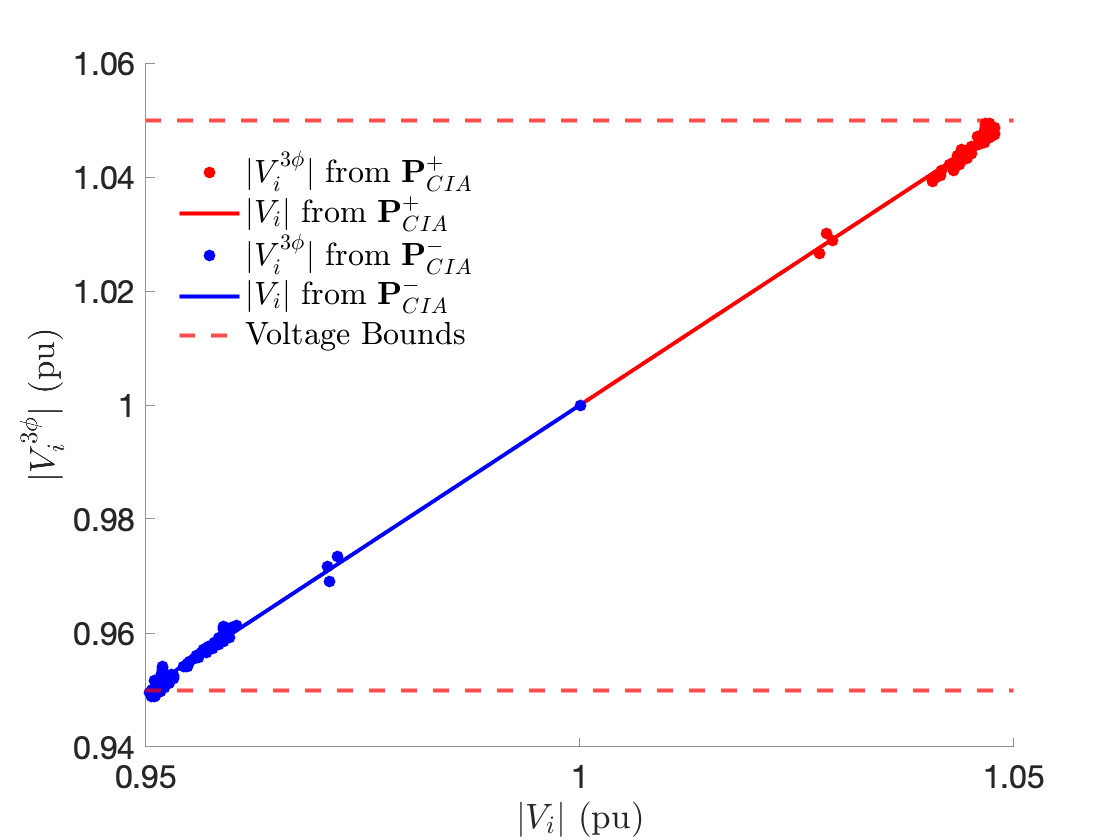}
\caption{Voltages magnitudes  for the IEEE 37-node system after modifying the impedance matrix based on Mod-Z approach.}
\label{fig:addedZ}
\end{figure}

In Fig.~\ref{fig:Base}, a scatter plot of predicted single-phase and actual three-phase voltage magnitudes is provided, i.e., $|V_i|$ vs. $|V^{3\phi}_i|$ for IEEE 37-node system, under Method~2ii. The red dots represent $|V^{3\phi}_i|$ with additional injections $p^+_{3\phi}$ and the blue dots correspond to $|V^{3\phi}_i|$ when demands $p^-_{3\phi}$ added. Fig.~\ref{fig:addedZ} shows the results after applying \textbf{Mod-Z}. Specifically, in $\mathbf{P}_\text{CIA}^{+}$ and $\mathbf{P}_\text{CIA}^{-}$, the impedance of each line is augmented by the mutual impedance from~\eqref{eq:zm_avg}.
As expected, using Mod-Z, the single-phase voltages closely approximate the three-phase voltages since mutual impedance is considered explicitly. The small differences in Fig.~\ref{fig:addedZ} between $|V^{3\phi}_i|$  and $|V_i|$ are caused by the approximation of $z_{ij}^m$ in~\eqref{eq:zm_avg}. 

It should be noted that using the modified impedance in $\mathbf{P}_\text{CIA}^{+}$ and $\mathbf{P}_\text{CIA}^{-}$ successfully increases the HC by incorporating the mutual impedances in the $\mathbf{P}_\text{CIA}^{+}$ problem. Specifically, $\overline{\text{HC}}$ increases from 25.09 MW to 30.22 MW (a 20\% increase), while $\underline{\text{HC}}$ increases from -14.89 MW to -19.30 MW (a 30\% increase). However, this increase can lead to (minor) voltage violations as seen in Fig.~\ref{fig:addedZ}.

To eliminate these (minor) violations, we present a simulation-based approach that selectively adjusts the impedance matrix in Mod-Z. Thus, instead of (naively) altering the impedance for all branches at once, only the branches connected to nodes with simulated voltage violations are modified. Specifically, we only modify the impedance of lines connected to nodes if they satisfy the following condition:
\begin{align}
    \left||V^{3\phi}_i|-|V_i|\right|>\epsilon\quad \forall i,
\end{align}
where $\epsilon$ is a design parameter that allows us to limit the number of line modifications. We denote Mod-Z($\epsilon$) as the Mod-Z method with the chosen parameter $\epsilon$. Modifying more lines leads to higher HC, but it comes at the cost of more voltage violations. No free lunch in engineering.

To explore this tradeoff further, Table~\ref{t:incrementalZ} tabulates the effects of different $\epsilon$ values in the \textbf{Mod-Z} approach. Clearly, with $\epsilon=0.0010$ pu, only 9 of 36 lines are modified in $\mathbf{P}_\text{CIA}^{+}$, while all voltage violations are eliminated, and the reduction in HC is less than 10\%. 


\begin{table}[t]
    \centering
    \caption{The impact of modifying the impedance matrix on HC and voltage violations for the modified IEEE 37-node system.
    }
    \label{t:incrementalZ}
    \begin{tabular}{p{2.5cm}p{1cm}p{1cm}p{1cm}p{1cm}}
        \toprule
        $\epsilon$ (pu) &0& 0.0005&0.0010&Method~2ii \\
        \midrule
        $\overline{\text{HC}}$ (MW) & 30.2 &27.5&27.4  &  25.09\\
        $\underline{\text{HC}}$ (MW) & -19.3 &   -17.5 &-17.3 & -14.9  \\
        $N_v$ & 10 & 5 & 0 & 0 \\
        $M_v$ & 0.0012 & 0.0004 & 0 & 0 \\
        \# modified lines in $\mathbf{P}_\text{CIA}^{+}$/$\mathbf{P}_\text{CIA}^{-}$ & 36/36 & 22/27 & 9/16 & - \\
        \bottomrule
    \end{tabular}
\end{table}

This section showed the value of selectively modifying line impedances to enable per-phase optimization to apply directly to unbalanced distribution systems. Next, we seek to further enlarge the three-phase HC by not just modifying impedances of each phase, but also by (incrementally) relaxing voltage bounds in the per-phase optimization formulation.   

\section{Iterative voltage bounds to increase HC}\label{sec:DHC}

In this section, a novel approach is introduced that incrementally improves the three-phase HC estimate by coupling 3-phase load flows with per-phase optimization $\mathbf{P}_\text{CIA}^{\phi,+\text{\textbackslash} -}$.  The proposed method is summarized in Fig.~\ref{fig:flowchart} and outlined as follows for $p^+$ $\overline{\text{HC}}$ (the approach is similar for $p^-$ and $\underline{\text{HC}}$): 
\begin{figure}[t]
\centering
\includegraphics[width=1\columnwidth]{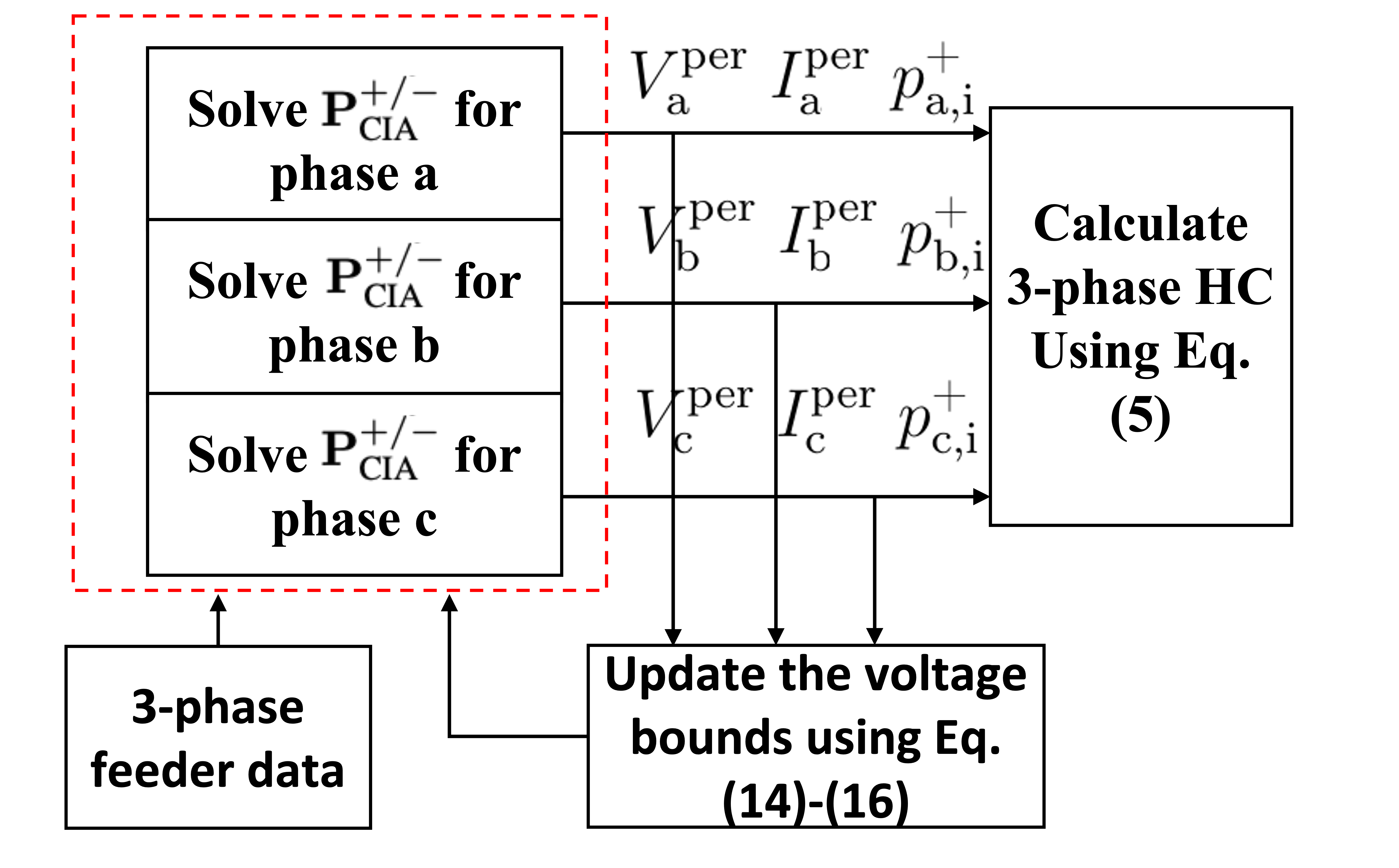}
\caption{Flowchart of the proposed iterative voltage bound approach. 
}
\label{fig:flowchart}
\end{figure}


\begin{enumerate}[label=\textbf{Step}~\arabic*:, leftmargin=1.25cm]
    \item \textit{Single-phase optimization}: given per-phase voltage bounds $\underline{V}$ and $\overline{V}$, solve $\mathbf{P}_\text{CIA}^+$ for each phase using Method 2ii to get nodal HC values $p_\phi^+$ and $\overline{\text{HC}}_\phi$. 
    
    \item \textit{Single-phase load flow}: Apply $p_\phi^+$ to each phase $\phi$ and perform single-phase load flow: $V^{\text{per}}_\phi$ and $I^{\text{per}}_\phi$. 
    

   \item \textit{Three-phase load flow}: Apply $\{p_\phi^+\}_{\phi=\{a,b,c\}}$ to 3-phase system and perform load flow: $V^{3\phi}$ and $I^{3\phi}$. 
   
    \item \textit{Termination condition}: The algorithm stops if any element of $|V^{3\phi}_{i}|$ exceeds $[\underline{V},\overline{V}]$.  
   
    \item    \textit{Estimate per-phase voltage}: The per-phase model ignores mutual impedances, which leads to a voltage difference across phases relative to the three-phase model.
    To estimate this difference, consider~\eqref{eq:volt_drop} and assume currents $I_{ij}^{\phi} \approx I_{ij,\phi}^{3\phi}$ are common across both the per-phase and three-phase systems. Then, the estimated voltage for each phase becomes,

    
    

    \begin{align}\label{eq:voltage_est}
         V_{j}^{\text{est}} = V_{i}^{\text{est}} -\begin{bmatrix}
    z^{\text{a}}_{ij} & z^{\text{ab}}_{ij}  & z^{\text{ac}}_{ij}  \\
    z^{\text{ba}}_{ij}  & z^{\text{b}}_{ij}  & z^{\text{bc}}_{ij}  \\
    z^{\text{ca}}_{ij}  & z^{\text{cb}}_{ij}  & z^{\text{c}}_{ij}  \\
    \end{bmatrix}I_{ij}^{3\phi} \quad \forall (i,j)\in \mathcal{E}.
    \end{align}
    
Since $v_0^{\text{est}}$, i.e. head node voltage is known, the voltage of other nodes of a radial grid can be found using~\eqref{eq:voltage_est}. 
   \item \textit{Per-phase voltage difference }: Using~\eqref{eq:voltage_est}, the difference in per-phase voltage can be found as,

       \begin{align}
        \Delta V_i = |V_{i}^{\text{est}}| -|V_i^{\text{per}}| \quad \forall i\in \mathcal{V}.
    \end{align}


   \item \textit{Updating voltage bounds}: the voltage bounds are updated for $\mathbf{P}_\text{CIA}^{\phi,+}$ to reflect the cumulative path voltage difference that arises due to per-phase optimization neglecting mutual impedances. The update is as follows:
    \begin{align}\label{eq:updateBound}
        \nonumber 
        &\overline{V} \leftarrow \overline{V} + \alpha \Delta V_i \\
        &\underline{V} \leftarrow  \underline{V} - \alpha \Delta V_i,
    \end{align}
    where $\alpha$ is a design parameter that can be set to less than 1 to allow smaller steps in each iteration. 
    \item \textit{Iterate}: Go to \textbf{Step}~1.
     
\end{enumerate}

Next section, numerical results are presented to validate the proposed methodologies. 

\section{Numerical results}\label{sec:numerical}
\begin{figure}[t]
\centering
\includegraphics[width=1\columnwidth]{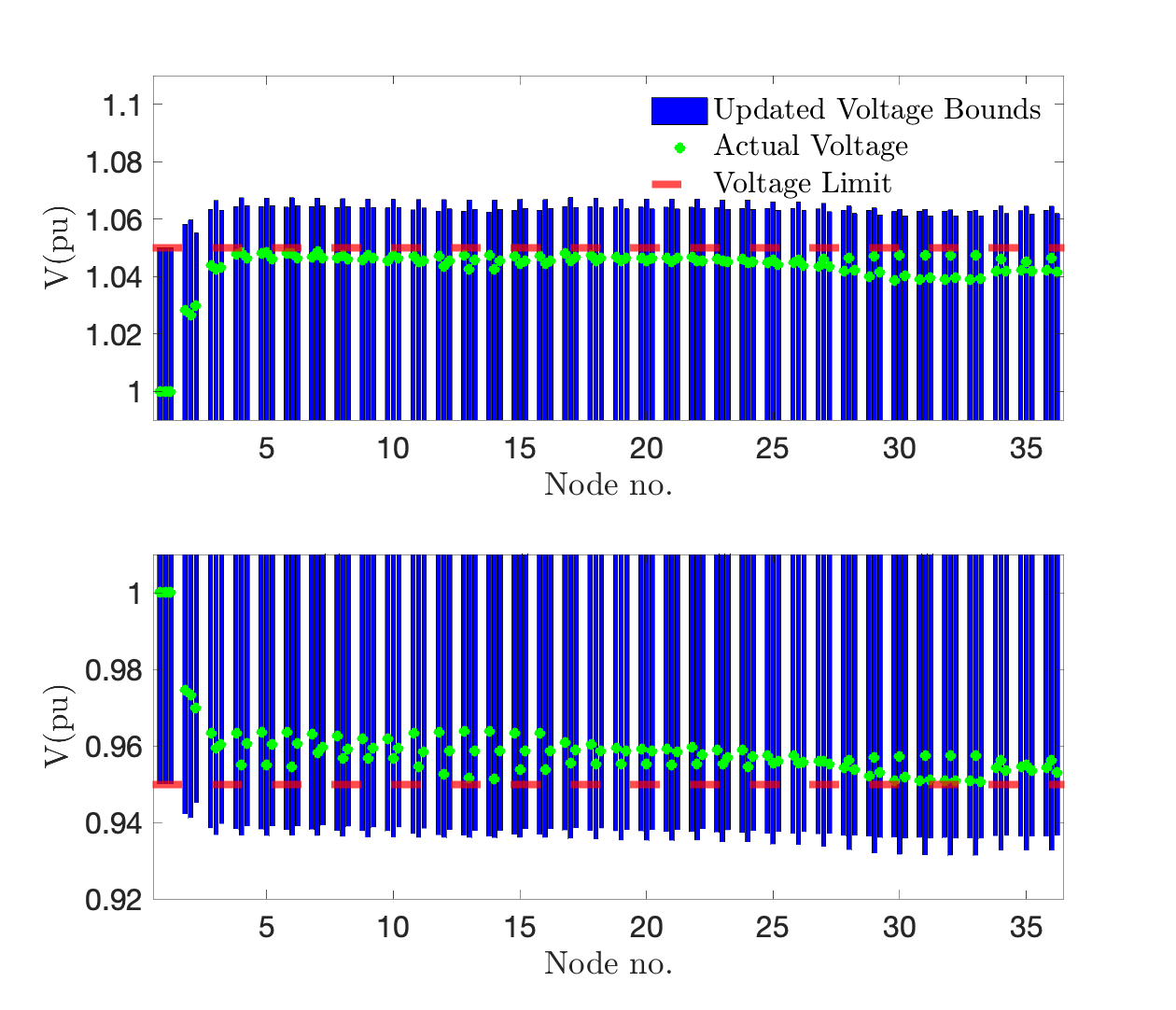}
\caption{Voltage bounds upon termination of the iterative method.
}
\label{fig:BoundDHC}
\end{figure}

In this section, simulation results on the IEEE 37-node test system are presented together with a realistic 534-node radial distribution system from Vermont. IEEE 37-node test system is a three-phase, unbalanced medium voltage (4.8 kV) network with a total load of 2.45 MW. The realistic feeder used in this paper is a 7.2 kV radial network including 534 nodes, 533 lines, and 160 loads with a total load of 2.47 MW. The MATLAB code provided by~\cite{garces2023matlab} is used for three-phase simulations. Using the proposed approach enables an increase in the amount of HC without causing any additional violations. Fig.~\ref{fig:BoundDHC} shows the voltage bounds upon the termination of the proposed iterative method. It can be seen that all of the three-phase voltages are within $\underline{V},\overline{V}$.

\begin{figure}[t]
\centering
\includegraphics[width=1\columnwidth]{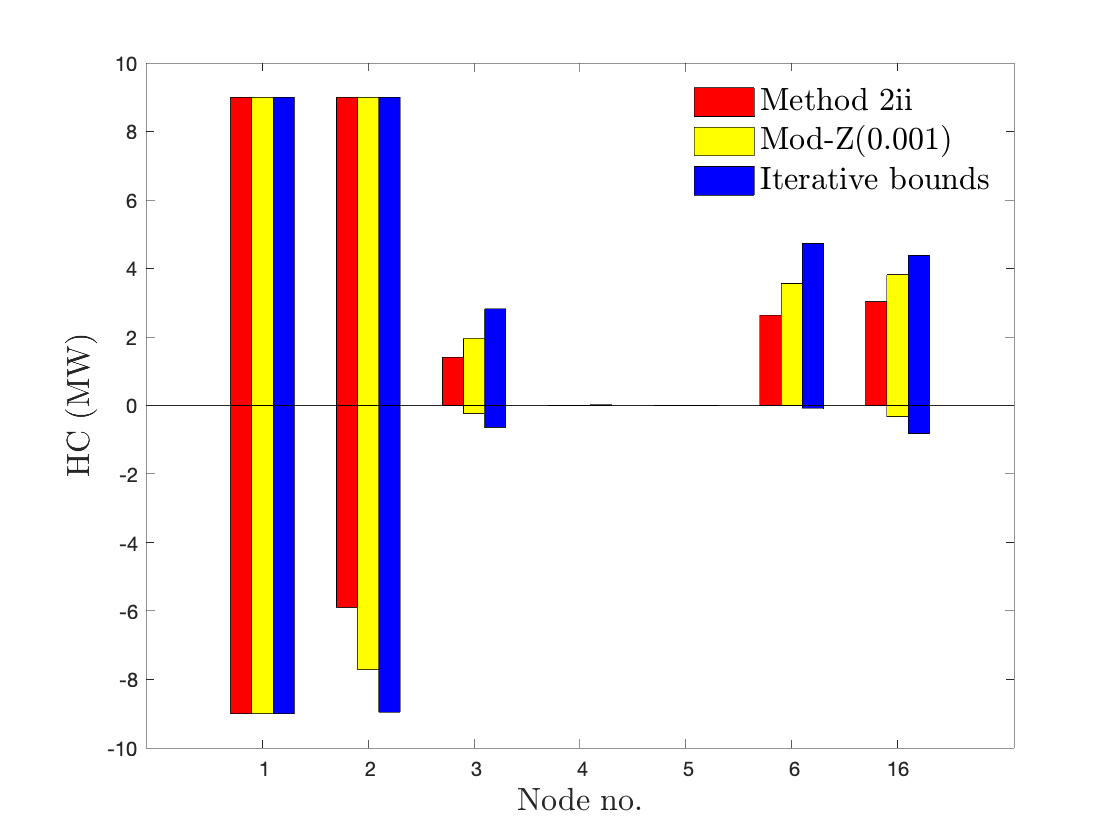}
\caption{Comparing the hosting capacity from the iterative approach to that of Method~2ii and Mod-Z.}

\label{fig:PDHC}
\end{figure}

For the IEEE 37-node system, the results of $p^-_i$ and $p^+_i$ obtained from three different methods—Method 2ii, Mod-Z(0.001), and the iterative HC approach—are displayed in Fig.~\ref{fig:PDHC}. Table~\ref{t:simulation_results} compares the simulation time and total HC for Method 2ii, Mod-Z(0.001), and the iterative method across two networks. It is worth noting that in the iterative method, $\underline{\text{HC}}$ and $\overline{\text{HC}}$ consistently show improvements when utilizing the iterative method. This enhancement is achieved by leveraging information regarding the mutual impedance of the grid. This increase in hosting capacity does not lead to any voltage violations, therefore no line modification is required in $\mathbf{P}_\text{CIA}^{\phi,+}$. That is, Mod-Z is not used with the iterative method. 


\begin{table}[htbp]
    \centering
    \caption{Comparing the different methods across two networks.}
    \label{t:simulation_results}
    \resizebox{\columnwidth}{!}{%
    \begin{tabular}{lcccccc}
        \toprule
        & \multicolumn{3}{c}{\textbf{IEEE 37 Node}} & \multicolumn{3}{c}{\textbf{534-node Feeder}} \\
        \cmidrule(lr){2-4} \cmidrule(lr){5-7}
        \textbf{Method} & HC & HC & Run Time & HC & HC & Run Time \\
        & (MW) & (MW) & (sec) & (MW) & (MW) & (sec) \\
        \midrule
        \textbf{Method 2ii} & -14.9 & 25.1 & 62 & -26.4 & 46.5 & 380 \\
        \textbf{Iterative HC} & -19.5 & 30.4 & 314 & -59.4 & 73.0 & 2973 \\
        \textbf{Mod-Z(0.001)} & -17.3 & 27.4 & 60 & -74.3 & 71.8 & 439 \\
        \textbf{Random Search} & -5.9 & 12.2 & 343 & -32.9 & 76.5 & 2346 \\
        \bottomrule
    \end{tabular}%
    }
\end{table}

\begin{figure}[t]
 \centering
\includegraphics[width=1\columnwidth]{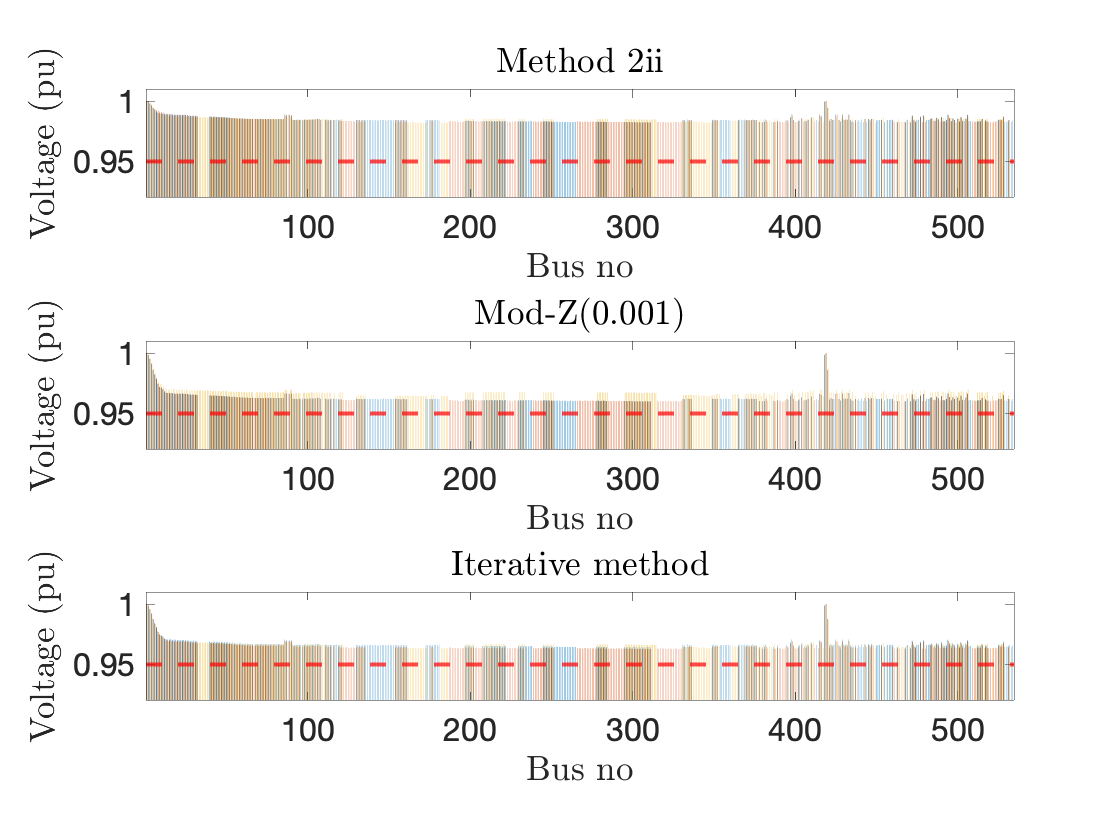}
\caption{Voltage profiles for a 534-node feeder are depicted in the figures below for $\mathbf{P}_\text{CIA}^{\phi,-}$. 
In these figures, {\color{blue}blue}, {\color{red}red}, and {\color{orange}yellow} correspond to phases {\color{blue}a}, {\color{red}b}, and {\color{orange}c}, respectively.}
\label{fig:534voltage}
\end{figure}
\begin{figure}[t]
 \centering
\includegraphics[width=1\columnwidth]{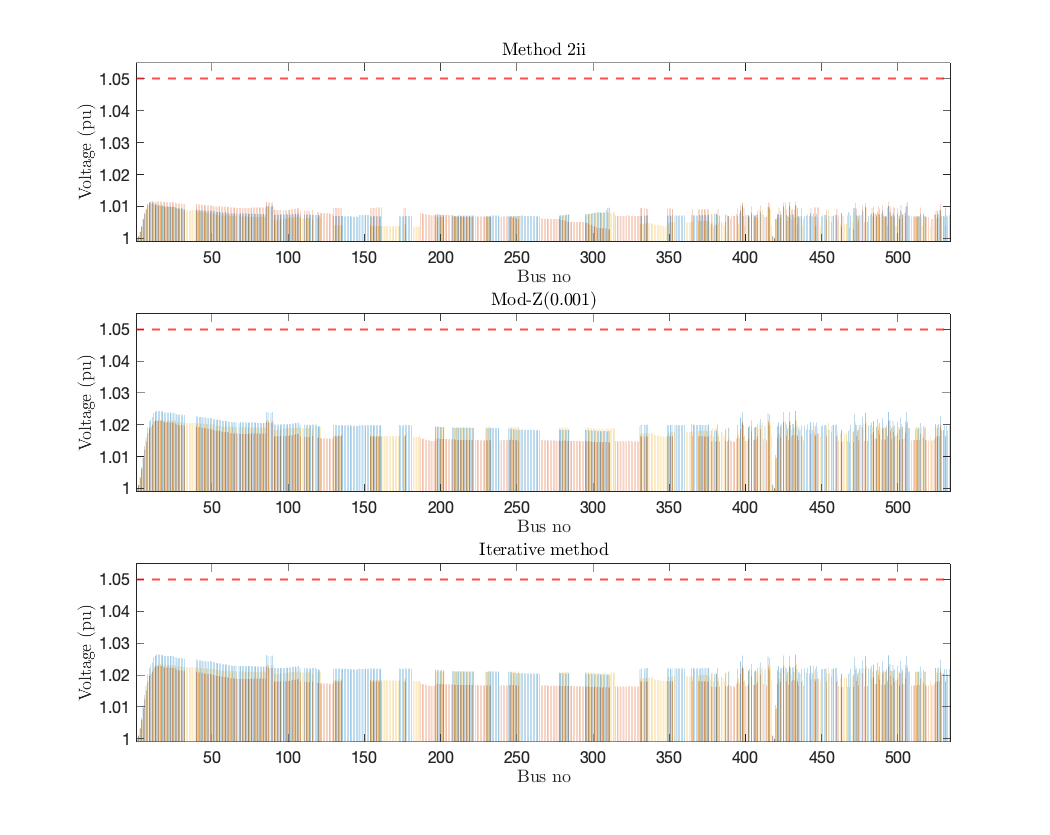}
\caption{Voltage profiles for a 534-node feeder are depicted in the figures below for $\mathbf{P}_\text{CIA}^{\phi,+}$. 
In these figures, {\color{blue}blue}, {\color{red}red}, and {\color{orange}yellow} correspond to phases {\color{blue}a}, {\color{red}b}, and {\color{orange}c}, respectively.}
\label{fig:534voltage-up}
\end{figure}

Figs.~\ref{fig:534voltage} and~\ref{fig:534voltage-up} present $|V^{3\phi}|$ for different methods applied to the 534-node network for $\mathbf{P}_\text{CIA}^{\phi,-}$ and $\mathbf{P}_\text{CIA}^{\phi,+}$. It can be seen that using the proposed iterative and Mod-Z methods, the voltage margin is smaller, which allows for higher HC as evident in Table~\ref{t:simulation_results}. In Figure \ref{fig:534voltage-up}, the voltages cannot approach the limits due to transformer rating constraints.

It is important to note that the optimization problem may result in very small HC values in some nodes while leading to significantly higher HC values in a few nodes within the system. This discrepancy can raise concerns regarding fairness since only certain consumers will be permitted to install DERs. We consider two sets of $w_i$, in the objective function and evaluate the HC for a realistic 534-node network under two scenarios: 1) Weight values ($w_i$) are equal for all nodes.
2) Weight values for leaf nodes are doubled compared to other nodes.
Modifying the $w_i$ coefficients enables us to expand the locations where DERs can be installed. Specifically, the locations with HC larger than 0.5 MW have increased from 8 to 16. However, this adjustment comes at the cost of reduced $\overline{\text{HC}}$, which decreases from 73.0 MW to 71.0 MW, and reduced $\underline{\text{HC}}$, which decreases from 59.4 MW to 53.0 MW.

Future research efforts could delve into exploring the trade-off between fairness in DER allocation and its impact on the overall HC of the grid.

\section{Conclusion}\label{sec:conclusion}

This paper has introduced a comprehensive approach to obtaining the DER HC in a three-phase distribution feeder. Leveraging CIA of the AC power flow, our methodology establishes bounds on positive and negative DER injections at each node. A rigorous analysis is developed to ascertain the conditions under which this per-phase approach can guarantee compliance with three-phase constraints. Furthermore, we have presented an iterative approach to enhance HC by adjusting per-phase voltage bounds. A simulation-based analysis using both the IEEE 37-node test feeder and a real 534-node unbalanced radial distribution feeder is performed and results demonstrate that the proposed iterative method increases the feeder HC. Potential future research encompasses the extension of the proposed method to analyze comprehensive 3-phase networks, as well as comparing its conservativeness to the method presented in this paper. Additionally, extending the HC analysis methods to meshed distribution and sub-transmission networks will be explored in future work.

\section*{Acknowledgments}
The authors appreciate Nawaf Nazir for his insightful discussions and support throughout the research. 

\appendix
\subsection{Derivation of current proxy bounds $l^-$ and $l^+$} \label{sec:appx}

The goal of this appendix is to clarify the structure of the affine  $f_\text{aff}(.)$ and quadratic $f_\text{quad}(.)$ functions that underpin bounds $l^-, l^+$ used in~\eqref{eq:proxies}. To derive the lower and upper bounds of $l$, we consider the second-order Taylor-series approximation of~\eqref{eq:current} about an appropriate nominal operating point, $x_{ij}^0 := \text{col}\{P_{ij}^0, Q_{ij}^0, v_j^0\}\in\mathbb{R}^3$. This yields an approximation that is accurate across a range of operating conditions~\cite{Nawaf2022TPWRS}: 
\begin{align}\label{eq:taylor}
        l_{ij}(P_{ij},Q_{ij},V_i) \approx l_{ij}^0(x_{ij}^0) + J_{ij}^\top\delta_{ij} + \frac{1}{2}\delta_{ij}^\top H_{e,ij}\delta_{ij},
\end{align}
where $\delta_{ij} := [P_{ij}-P_{ij}^0,Q_{ij}-Q_{ij}^0,v_{j}-v_{j}^0]$, the Jacobian, $J_{ij}$, and Hessian,  $H_{e,ij}$, are defined as
\begin{align}\label{eq:jacobian}
 J_{ij} := 
 \begin{bmatrix}
\frac{2P_{ij}^0}{v_i^0} & \frac{2Q_{ij}^0}{v_i^0} & -\frac{(P_{ij}^0)^2+(Q_{ij}^0)^2}{(v_i^0)^2}
\end{bmatrix},
\end{align}
\begin{align}
 H_{e,ij} := \begin{bmatrix}
     \frac{2}{v_i^0} & 0 & \frac{-2P_{ij}^0}{(v_i^0)^2}\\
     0 & \frac{2}{v_i^0} & \frac{-2Q_{ij}^0}{(v_i^0)^2}\\
     \frac{-2P_{ij}^0}{(v_i^0)^2} & \frac{-2Q_{ij}^0}{(v_i^0)^2} & 2\frac{(P_{ij}^0)^2+(Q_{ij}^0)^2}{(v_i^0)^3}
 \end{bmatrix}. \label{eq:hessian}
\end{align}

From~\eqref{eq:taylor}, the square of current magnitude is always positive, so:
\begin{align}\label{eq:l_est}
    l_{ij} = \left|l_{ij}\right|\approx \left|l_{ij}^0 + J_{ij}^\top\delta_{ij} + \frac{1}{2}\delta_{ij}^\top H_{e,ij}\delta_{ij}\right|.
\end{align}

Applying the triangle inequality and the fact that Hessian in~\eqref{eq:hessian} is positive semi-definite (PSD)~\cite{nazir2021TPWRS}, we have
\begin{align}\label{eq:triangular}
      l_{ij} 
      \le l_{ij}^0 + \left|J_{ij}^\top\delta_{ij}\right| + \frac{1}{2}\delta_{ij}^\top H_{e,ij}\delta_{ij}.
\end{align}


Applying the properties of the maximum operator, we get the quadratic function:
\begin{align}\label{eq:max}
   l_{ij} \le l^0_{ij} +  \max \left\{2\left|J_{ij}^\top\delta_{ij}\right|, \delta_{ij}^\top H_{e,ij}\delta_{ij}\right\}.
\end{align}
Note that the RHS of~\eqref{eq:max} is quadratic in terms of the three physical variables $(P_{ij},Q_{ij},V_i)$ that embody $\delta_{ij}$. To characterize the upper bound in terms of the proxy variables requires considering worst-case combinations of upper ($^+$) and lower ($^-$) proxy variables, i.e., over all eight combinations: $\delta_{ij}^+:=\delta_{ij}(P_{ij}^+,Q_{ij}^+,V_i^+)$, $\delta_{ij}(P_{ij}^+,Q_{ij}^+,V_i^-)$, $\hdots$, $\delta_{ij}(P_{ij}^-,Q_{ij}^-,V_i^+)$,  and $\delta_{ij}^-:=\delta_{ij}(P_{ij}^-,Q_{ij}^-,V_i^-)$. Thus, we get: 
\begin{align}
f_\text{quad}(.) := l_{ij}^0 + \max \left\{2\left|J_{ij,+}^\top \delta_{ij}^+ + J_{ij,-}^\top \delta_{ij}^-\right|,\psi_{ij}\right\},    
\end{align}
where $J_{ij,+}$ and $J_{ij,-}$ are composed of the positive and negative entries of $J_{ij}$, respecitively, and $J_{ij} = J_{ij,+} + J_{ij,-}$. Further, $\psi_{ij} := \max\{\delta_{ij}^{+/-} H_{\text{e},ij} \delta_{ij}^{+/-} \}$ is the largest product among the eight proxy pairs.
Clearly, relaxing $f_\text{quad}(.)$ provides a convex upper bound on $l_{ij}$ as utilized in~\eqref{eq:proxies}.

For the lower bound, consider~\eqref{eq:taylor} and drop the  term with PSD $H_{\text{e},ij}$, which gives
\begin{align}
    l_{ij} \ge l_{ij}^0 + J_{ij}^\top \delta_{ij} := \underline{l}_{ij}.
\end{align}
Thus, in terms of proxy variables, we get 
\begin{align}
    f_\text{aff}(.) := l_{ij}^0 + J_{ij,+}^\top \delta_{ij}^- + J_{ij,-}^\top \delta_{ij}^+.
\end{align}
This completes the derivations. For full details on these bounds and the CIA-based methods and results (for balanced feeders), please see~\cite{nazir2021TPWRS,Nawaf2022TPWRS}. 

\subsection{Proof of Theorem~\ref{thm:1}} \label{appx:proof}
From Assumption~\ref{ass:imp}, the impedance matrix has identical mutual impedances $z_{ij}^m$, which together with~\eqref{eq:z_line}, means that $\Delta V^{3\phi}_{ij}$ can be expressed as,
\begin{align}
\Delta V^{3\phi}_{ij}=\begin{bmatrix}
z^{a}_{ij}I^a_{ij}+z^{m}_{ij}(I^b_{ij}+I^c_{ij})     \\
z^{b}_{ij}I^b_{ij}+z^{m}_{ij}(I^a_{ij}+I^c_{ij})     \\
z^{c}_{ij}I^c_{ij}+z^{m}_{ij}(I^a_{ij}+I^b_{ij})  
\end{bmatrix}.
\end{align}

Now, under Assumption~\ref{ass:balancedLoad}, $I_{ij}^a+I_{ij}^b+I_{ij}^c = 0$, which decouples the phases as
\begin{align}\label{eq:ohmLaw}
\Delta V^{3\phi}_{ij}=\begin{bmatrix}
z_{ij}^a-z_{ij}^m   & 0   & 0   \\
0    & z_{ij}^b-z_{ij}^m  & 0    \\
0  & 0    & z_{ij}^c-z_{ij}^m 
\end{bmatrix}I_{ij}^{3\phi}.
\end{align}
The diagonal structure clearly extends per-phase analyses to the corresponding full 3-phase (unbalanced) feeder. Thus, HC analysis via $\mathbf{P}_\text{CIA}^{\phi,+}$ meets 3-phase voltage requirements. This concludes the proof.


\bibliographystyle{IEEEtran}
\bibliography{ref}

\end{document}